\documentclass[12pt]{article}
\usepackage{hyperref,graphicx}

\def\ba{\begin{eqnarray}}
\def\ea{\end{eqnarray}}
\def\bq{\begin{eqnarray*}}
\def\eq{\end{eqnarray*}}
\def\be{\begin{equation}}
\def\ee{\end{equation}}
\def\bm{\begin{math}}
\def\me{\end{math}}
\def\bi{\bibitem}

\def\la{\langle}
\def\ra{\rangle}

\def\prl#1#2#3{{Phys. Rev. Lett.} \textbf{#1}, #2 (#3)}

\def\pra#1#2#3{Phys. Rev. A \textbf{#1}, #2 (#3)}
\def\prb#1#2#3{Phys. Rev. B \textbf{#1}, #2 (#3)}
\def\pre#1#2#3{Phys. Rev. E \textbf{#1}, #2 (#3)}

\begin{document}

\begin{center}

{\Large \bf Phase Separation Driven by Surface Diffusion: A Monte Carlo Study}
\vskip1cm
S. van Gemmert$^1$, G.T. Barkema$^1$ and Sanjay Puri$^2$ \\
\vskip0.25cm
{\it $^1$Institute for Theoretical Physics \\
Utrecht University, Leuvenlaan 4 \\
3584 CE Utrecht, THE NETHERLANDS.} \\
\vskip0.25cm
{\it $^2$School of Physical Sciences \\
Jawaharlal Nehru University \\
New Delhi -- 110067, INDIA.}
\end{center}

\vskip2cm
\begin{abstract}
We propose a kinetic Ising model to study phase separation driven by surface
diffusion. This model is referred to as \emph{Model S}, and consists of the
usual Kawasaki spin-exchange kinetics (\emph{Model B}) in conjunction with
a kinetic constraint. We use novel multi-spin coding techniques to develop
fast algorithms for Monte Carlo simulations of Models B and S. We use these
algorithms to study the late stages of pattern dynamics in these systems.
\end{abstract}

\newpage
\section{Introduction}

Consider a homogeneous binary (AB) mixture, which is rendered thermodynamically
unstable by a rapid temperature quench below the miscibility gap.
The system prefers to be in a phase-separated state at the lower temperature.
The far-from-equilibrium evolution
of the system from the unstable homogeneous state to the segregated
state has received considerable attention \cite{ab94,bf01,ao02,dp03}.
This evolution is characterized by the emergence and growth of domains enriched
in the components A and B. The terms used to describe this nonequilibrium
process are {\it phase ordering dynamics}, {\it domain growth} or {\it coarsening}.
A quantitative characterization of phase ordering systems focuses on
(a) the domain growth law; (b) the statistical properties of the evolution morphology;
and (c) the temporal correlation of pattern dynamics.

The equilibrium phase-separated state
is uniquely determined by its thermodynamic properties.
However, there is a diverse range of kinetic pathways which enable segregation.
For example, phase separation in alloys is usually driven by vacancy-mediated
diffusion \cite{sp97}. On the other hand, for fluid mixtures, hydrodynamic
velocity fields enable convective transport of material along domain
boundaries and give rise
to novel asymptotic behaviors \cite{es79}. Furthermore, phase separation in
mixtures can be frozen (or near-frozen) into mesoscopic states
by the presence of quenched disorder \cite{pcp91}, viscoelastic effects
\cite{op99,ao02}, etc.

In this paper, we present results from a comparative Monte Carlo (MC) study of two
kinetic Ising models for phase separation in
binary mixtures. The first of these is the usual Kawasaki spin-exchange model
\cite{kk72,dp03}, which mimics segregation via diffusion. The second model
mimics the case where only surface diffusion is permitted. An important
goal of this paper is methodological, viz., the formulation of a kinetic
Ising model where bulk diffusion is suppressed. Another important goal is to
compare pattern dynamics for phase separation with and without bulk diffusion.

This paper is organized as follows. In Sec.~2, we describe the kinetic Ising
models studied here and our MC simulation techniques.
Our MC approach uses novel multi-spin
coding techniques, which enable large-scale and long-time simulations of these
models. In Sec.~3, we discuss the domain growth laws which
arise from bulk and surface diffusion, and also present detailed numerical results.
Finally, Sec.~4 concludes this paper with a summary and discussion of
our results.

\section{Numerical Methodology}

\subsection{Kinetic Ising Models}

The standard model for binary mixtures is the Ising model:
\ba
\label{ham}
{\cal H}= -J\sum_{\langle ij \rangle} \sigma_i \sigma_j , \quad \sigma_i = \pm 1,
\ea
where the spins $\{\sigma_i\}~(i=1 \rightarrow N)$ are located on
a discrete lattice. The states $\sigma_i = +1$ or $-1$ denote the presence of an
A-atom or B-atom at site $i$, respectively. We consider the case with
ferromagnetic ($J>0$) nearest-neighbor interactions, denoted by the subscript
$\langle ij \rangle$ in Eq.~(\ref{ham}). The phase diagram for the binary
mixture is obtained in an ensemble with fixed temperature
$T$ and magnetization $M = \sum_i \sigma_i$.

The Ising system does not have an intrinsic dynamics as the Poisson brackets
(or commutators) of spin variables are identically zero. Therefore, one
introduces stochastic kinetics by placing the system in contact with
a heat-bath which induces fluctuations. The Ising model, in conjunction with
a physically appropriate spin kinetics, is referred to as a \emph{kinetic
Ising model} \cite{dp03,bh02}. An important example is the
Kawasaki spin-exchange model \cite{kk72}, which has nearest-neighbor spin
exchanges with Metropolis acceptance probabilities. In an MC simulation
of this model, a pair of nearest-neighbor sites $i$ and
$j$ is randomly selected, and the spins $\sigma_i$ and $\sigma_j$
are exchanged. The probability that this exchange is accepted is given by
\ba
\label{prob}
P &=& \min \left[ 1, \exp(-\beta \Delta {\cal H}) \right], \nonumber \\
\Delta {\cal H} &=& J(\sigma_i - \sigma_j) \left( \sum_{L_i \neq j} \sigma_{L_i}
- \sum_{L_j \neq i} \sigma_{L_j} \right) .
\ea
Here, $\Delta {\cal H}$ is the energy change due to the proposed
spin exchange, and $\beta=(k_BT)^{-1}$ is the inverse temperature, with
$k_B$ denoting the Boltzmann constant. In Eq.~(\ref{prob}), $L_i$ denotes
the nearest-neighbors of $i$ on the lattice.
A single Monte Carlo step (MCS) corresponds
to $N$ such attempted exchanges. A large number of
MC simulations of the Kawasaki model have been reported in the literature
\cite{asm88,mb95}.

The phase-separation kinetics in this microscopic model is analogous to that
for the coarse-grained Cahn-Hilliard-Cook (CHC) equation,
which is obtained as follows:
\ba
\label{chc}
\frac{\partial}{\partial t} \psi (\vec{r},t) &=& - \vec{\nabla} \cdot
{\vec{J}}(\vec{r},t) \nonumber \\
&=& \vec{\nabla} \cdot \left[ D\vec{\nabla} \mu (\vec{r},t) + \vec{\theta} (\vec{r},t)
\right] \nonumber \\
&=& \vec{\nabla} \cdot \left[ D\vec{\nabla} \left( \frac{\delta \cal{F}}{\delta \psi}
\right) + \vec{\theta} (\vec{r},t) \right] .
\ea
Here, $\psi (\vec{r},t)$ is the order parameter at space point
$\vec{r}$ and time $t$. Typically, $\psi (\vec{r},t)=\rho_A (\vec{r},t)-\rho_B
(\vec{r},t)$, where $\rho_A$ and $\rho_B$ denote the local densities of
species A and B. In Eq.~(\ref{chc}), the quantities $\vec{J}, D$ and $\mu$ denote the
current, diffusion coefficient, and chemical-potential difference between A and B,
respectively. The chemical potential is obtained as a functional derivative of the
Helmholtz free energy, which is often taken to have the $\psi^4$-form:
\ba
\label{psi4}
{\cal F} [\psi] &=& {\cal H} - TS \nonumber \\
&\simeq& \int d\vec{r} \left[ -\frac{1}{2} k_B (T_c-T) \psi^2 +
\frac{k_BT_c}{12} \psi^4 + \frac{J}{2} (\vec{\nabla} \psi)^2  \right],
\ea
where we have identified $\la \sigma_i \ra = \psi (\vec{r}_i)$ in Eq.~(\ref{ham}) and
Taylor-expanded various terms. Here, $T_c$ denotes the critical temperature.
Finally, the Gaussian white noise term $\vec{\theta} (\vec{r},t)$
in Eq.~(\ref{chc}) has zero average
and obeys the appropriate fluctuation-dissipation relation. The CHC equation is
also known as {\it Model B} in the Hohenberg-Halperin classification scheme
for critical dynamics \cite{hh77}. Further, using a master-equation approach, the
CHC equation can be motivated from the spin-exchange model \cite{kb74}. Therefore,
we will subsequently refer to the Kawasaki model as ``Model B''.

Before proceeding, we stress that the general form of the CHC equation
contains an order-parameter-dependent mobility \cite{lbm75,ki78,pbd92}:
\ba
\label{opm}
D(\psi) = D_0 \left( 1 - \frac{\psi^2}{{\psi_0}^2} \right) ,
\ea
where $\psi_0$ is the saturation value of the order parameter at $T=0$. This
is not consequential for quenches to moderate temperatures, but plays an
important role for deep quenches where $\psi \simeq \pm \psi_0$ in bulk domains.
In that case, bulk diffusion is effectively eliminated and domain growth
proceeds by surface diffusion \cite{pbl97,cy89,lm92}. In the context of the
Kawasaki model, this can be understood by focusing on an interfacial pair
with the minimum barrier for interchange: $\sigma_i = +1$ at the periphery of
an up-rich domain and having only one neighbor with the same spin value,
and $\sigma_j = -1$ in a down-rich domain. At low temperatures, the bulk domains
are very pure and the energy barrier to the interchange $\sigma_i \leftrightarrow
\sigma_j$ is $\Delta {\cal H} = 4J$. Thus,
the time-scale for this interchange $\tau_K \sim \exp (\beta \Delta {\cal H})
\rightarrow \infty$ as $T \rightarrow 0$, effectively blocking bulk diffusion.
Of course, once an impurity spin is placed inside a bulk domain, there is
no further barrier to its diffusion.

Apart from this natural blocking of bulk diffusion at $T=0$, there are systems
where the bulk mobility diminishes drastically due to physical processes,
e.g., one or both of the components may undergo a glass \cite{sj97} or
gelation \cite{sb93,sp01} transition. At the phenomenological level, this
has been modeled by setting the mobility to zero in regions rich in the
glass-phase or gel-phase. At the microscopic level, we propose
a kinetic Ising model where bulk diffusion is suppressed by introducing
a kinetic constraint. We disallow exchanges $\sigma_i \leftrightarrow
\sigma_j$ if the neighboring spins of the pair are
all parallel, even though such an exchange would not raise the energy.
In this case, segregation is driven primarily by diffusion along domain
boundaries, though some bulk transport occurs via impurity $n$-spin clusters.
(This bulk diffusion is negligible for moderate to deep quenches.) We will
subsequently refer to this model as ``Model S'' \cite{pbl97}. Clearly, Model S
can be generalized to the case of reduced (though non-zero) mobility in the bulk
domains. This is done by allowing spin-exchanges with different time-scales
depending on the number and type of parallel neighbors for a spin pair.

\subsection{Numerical Details}

All our MC simulations were performed on an $L \times L$ square lattice
with periodic boundary conditions. At time $t=0$, the temperature
was quenched from $T=\infty$ to $T<T_c$, where
$T_c \simeq 2.269$ is the critical temperature of the $d=2$ Ising model.
(All energy scales are measured in units of $J$, and we set the
Boltzmann constant $k_B = 1$.) The disordered
initial state consisted of a uniform mixture of $N_A$ A-atoms
and $N_B$ B-atoms with $N=N_A+N_B$. The case with $N_A=N_B$ corresponds
to a critical quench.

Our MC simulations exploit the technique of \emph{multi-spin coding}.
For a general introduction to this technique, see Ref.~\cite{nb99}.
The basic idea is that one can exploit the 64-bit computer architecture
to undertake a parallel simulation of 64 systems. This is done by storing
the spin $\sigma_i$ at site $i$ in the $k^{\rm{th}}$ system
in the $k^{\rm{th}}$ bit of a 64-bit word $S[i]$. Recall that, in
one elementary move for Model B, we propose to exchange the spins located
on nearest-neighbor sites $i$ and $j$. For Model S, we impose the kinetic
constraint that a pair of spins surrounded by aligned neighbors is not exchanged.

For the $d=2$ square lattice considered here, each site has four
nearest-neighbors. Let $n_0$, $n_1$ and $n_2$ be the three nearest-neighbors
(other than $j$) of site $i$. Similarly, let $m_0$, $m_1$ and $m_2$ be the
three nearest-neighbors (other than $i$) of site $j$. To determine the change
in energy resulting from the proposed spin exchanges in all 64 simulations,
we first identify which of the six neighbors $(n_0,n_1,n_2,m_0,m_1,m_2)$
are antiparallel. This can be done in six operations with the {\it exclusive or}
operation $\oplus$:
\begin{eqnarray}
A_k &=& S[i] \oplus S[n_k] , \quad k=0 \rightarrow 2,\nonumber \\
B_k &=& S[j] \oplus S[m_k] , \quad k=0 \rightarrow 2.
\end{eqnarray}
The energy change associated with the spin exchange and (thereby) the
acceptance probability is governed by the number of antiparallel
spins.  In an ordinary program, this would involve a summation over the
surrounding spins. With logical operations, it is more convenient to
determine the logical variables $P_k$ which tell whether
$\sigma_i$ is antiparallel to at least $k$ of its neighbors (other than
$\sigma_j$). Note that the Metropolis algorithm only requires
$P_1, P_2$ and $P_3$. These can be obtained with six operations:
\begin{eqnarray}
P_2 &=& A_0 \land A_1 , \nonumber \\
P_1 &=& A_0 \lor  A_1 , \nonumber \\
P_3 &=& A_2 \land P_2 , \nonumber \\
P_2 &=& P_2 \lor ( A_2 \land P_1) , \nonumber \\
P_1 &=& P_1 \lor A_2 .
\end{eqnarray}
Similarly, the variables $Q_k$ that tell whether $\sigma_j$ is
antiparallel to at least $k$ of its neighbors (other than $\sigma_i$)
can be obtained with six operations.

Finally, the acceptance probability for the proposed spin exchanges is
obtained by using \emph{random bit patterns} $R_0$, $R_1$ and
$R_2$. These are designed so that the probability for each
bit to be 1 is $P_b=\exp(-4\beta J)$. Thus, the following statements
comprise the core of our Model S algorithm:
\begin{eqnarray}
\rm{Flip}  &=&  (S[i] \oplus S[j]) \land (P_1 \lor Q_3 \lor R_0) \land (P_2 \lor Q_2 \lor R_1)
\land (P_3 \lor Q_1 \lor R_2) , \nonumber \\
\rm{Flip}  &=&  \rm{Flip} \land (P_1 \lor \lnot Q_3)
\land (Q_1 \lor \lnot P_3) , \nonumber \\
S[i]  &=&  S[i] \oplus \rm{Flip} , \nonumber \\
S[j]  &=&  S[j] \oplus \rm{Flip} .
\end{eqnarray}
The implementation of Model B dynamics is simply obtained by omitting the
second of the above statements.

These 36 operations for Model S (or 30 for Model B) act on all 64 bits and thus
perform 64 elementary moves.  In conjunction with the required {\it load}
and {\it store} operations, and generation of the random
bit patterns, our implementation of Model B for a $512^2$ system
with multi-spin coding requires 4.6 ns CPU-time per elementary
move on an AMD-64 computer with 3 GHz clock frequency.
This should be contrasted with a direct implementation
of this model, which requires approximately 100 ns CPU-time
per elementary move on the same machine.

The procedure outlined above, which simulates 64 separate systems,
is known as a \emph{synchronous} multi-spin algorithm~\cite{nb99}.
The boundaries of these 64 systems can be glued together to yield an
\emph{asynchronous} multi-spin algorithm~\cite{nb99}, simulating one
system which is 64 times larger. This comes at the cost of (a) more
complicated programming; and (b) a small reduction in the program
efficiency. The statistical results for domain morphologies
presented in Secs.~3.1, 3.2 and 3.3 were obtained by averaging over 150
asynchronous simulations with system sizes $L=512$. The results for the
autocorrelation function in Sec.~3.4 were obtained by averaging over 64
synchronously simulated systems with $L=1024$.

\section{Detailed Results}

\begin{figure}[htb]
\centering
\includegraphics[width=0.6\textwidth]{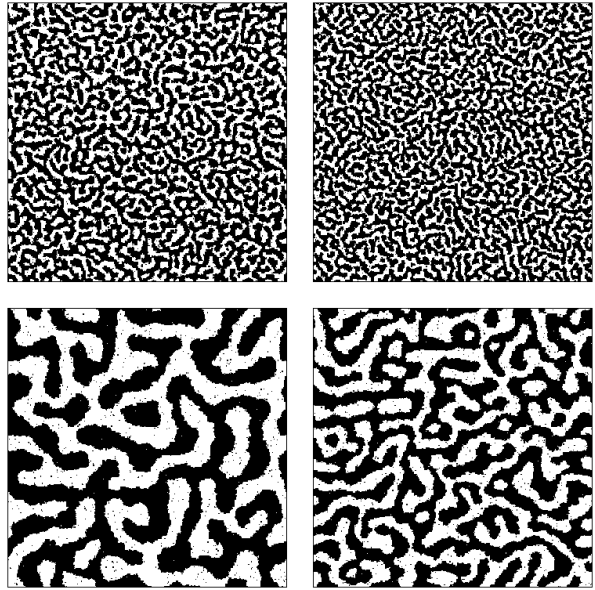}
\caption{Evolution pictures for phase separation in a binary (AB) mixture with a
critical composition. The component A ($\sigma = +1$) is marked in black,
and the component B ($\sigma = -1$) is unmarked. The system was quenched
at time $t=0$ from $T=\infty$ to $T=0.63T_c$. The top and bottom panels show
snapshots at times $t=10^4$ and $10^6$ MCS, using either Model
B dynamics (left), or Model S dynamics (right). The MC simulations were
done on square lattices of size $512^2$ with periodic boundary conditions.
The details of the simulations are described in the text.
\label{snap}}
\end{figure}

As stated earlier, the initial condition for our MC simulations consists
of a random configuration. The temperature is
quenched to $T<T_c$ at $t=0$, and the system evolves via either Model B or Model
S dynamics towards its new equilibrium state. Figure~\ref{snap} shows the typical
time evolution for a critical composition (50~\% A and 50~\% B)
after a quench to $T=0.63T_c$ for Model B (left) and Model S
(right). Notice that the evolution morphology has a characteristic
domain size, which we denote as $R(t)$.
The growth process is substantially slower for S-dynamics,
as expected. We will demonstrate shortly that the growth law due to
bulk diffusion is $R(t) \sim t^{1/3}$, which is referred to as the
Lifshitz-Slyozov (LS) growth law. The corresponding law for segregation
via surface diffusion is $R(t) \sim t^{1/4}$.
However, we reiterate that bulk diffusion is
not eliminated entirely in Model S because of the presence of impurity
spin clusters in bulk domains. At high temperatures, there is
a reasonable fraction of impurity spins and we expect the S-dynamics
to cross over to $t^{1/3}$-growth at late times. The crossover time increases
rapidly at lower temperatures where there are very few impurity spins in
the bulk as $P_{\rm{imp}} \simeq \left[ 1 + \exp(8 \beta J) \right]^{-1}$
in $d=2$.

We will study the evolution depicted in Fig.~\ref{snap} using quantities like the
correlation function and autocorrelation function.

\subsection{Growth Laws}

The first relevant property is the growth law governing the
segregation process. We computed the typical domain size $R(t)$ as the
first zero-crossing of the two-point correlation function:
\begin{eqnarray}
\label{cor}
C(\vec{r},t) & = & {1 \over N} \sum_{i=1}^N \left[ \la \sigma_i (t)
\sigma_{i + \vec{r}} (t)\ra  - \la \sigma_i (t)\ra \la \sigma_{i + \vec{r}} (t)
\ra \right] \\
\label{ds}
& \equiv & g \left( {r \over R} \right) .
\end{eqnarray}
Here, $\vec{r}$ denotes the displacement vector, and we consider systems
which are translationally invariant and isotropic.
The angular brackets in Eq.~(\ref{cor}) denote an averaging over
independent initial conditions and noise realizations.
Equation~(\ref{ds}) is the dynamical-scaling property of the correlation
function \cite{bs74}, and reflects the fact that the morphology is
self-similar in time, upto a scale factor (see Fig.~\ref{snap}).
One can use other definitions of the length scale also, but these are
all equivalent in the scaling regime.

At this stage, it is useful to clarify the domain growth laws which arise
due to bulk and surface diffusion. A convenient starting point is
the CHC equation (\ref{chc}) with an order-parameter-dependent mobility
$D(\psi)$. We consider a general situation where
the diffusion coefficient at the interface
($\psi = 0$) is $D_s$, and that in the bulk [$\psi = \psi_s (T)$] is $D_b$
with $D_b \leq D_s$. This difference in surface and bulk mobilities may
be the consequence of low-temperature dynamics or due to physical processes
like glass formation or gelation.
Then, the simplest functional form which models the mobility is
\ba
D(\psi) = D_s \left( 1 - \alpha \frac{\psi^2}{\psi_s^2} \right),
\quad \alpha = 1 -\frac{D_b}{D_s} ,
\ea
which is equivalent to Eq.~(\ref{opm}) with $D_0 = D_s$ and $\psi_0^2 =
\psi_s^2/\alpha$. We focus on the deterministic case of Eq.~(\ref{chc}):
\ba
\label{ch}
\frac{\partial}{\partial t} \psi (\vec{r},t) = D_s \vec{\nabla} \cdot
\left\{ \left( 1 - \alpha \frac{\psi^2}{{\psi_s}^2} \right) \vec{\nabla}
\left[ -(T_c - T) \psi + \frac{T_c}{3} \psi^3 - J \nabla^2 \psi \right] \right\} ,
\ea
where we have used the $\psi^4$-form of the free energy from Eq.~(\ref{psi4}).
The saturation value of the order parameter in Eq.~(\ref{ch}) is
$\psi_s (T) = \sqrt{3 \left(1-T/T_c\right)}$. Using the natural scales for
the order parameter, length and time, we can rewrite Eq.~(\ref{ch})
in the dimensionless form:
\ba
\label{chd}
&& \frac{\partial}{\partial t} \psi (\vec{r},t) = \vec{\nabla} \cdot
\left[ \left( 1 - \alpha \psi^2 \right) \vec{\nabla}
\left( -\psi + \psi^3 - \nabla^2 \psi \right) \right] , \nonumber \\
&& \alpha \in [0,1]~~\mbox{for}~~ D_b \leq D_s .
\ea

The RHS of Eq.~(\ref{chd}) can be decomposed as \cite{lm92}
\ba
\label{chsplit}
\frac{\partial}{\partial t} \psi (\vec{r},t) &=&
(1-\alpha) \nabla^2 (-\psi + \psi^3 - \nabla^2 \psi) + \nonumber \\
&& \alpha \vec{\nabla} \cdot \left[ \left( 1 - \psi^2 \right) \vec{\nabla}
\left( -\psi + \psi^3 - \nabla^2 \psi \right) \right] ,
\ea
where the first term on the RHS corresponds to bulk diffusion. This term
disappears for $\alpha = 1$ or $D_b=0$. The second term on the RHS corresponds to
surface diffusion and is only operational at interfaces where $\psi \simeq 0$.
Following Ohta \cite{to88}, we can obtain an equation which describes
the interfacial dynamics. The location of the interfaces $\vec{r}_i(t)$ is
defined by the zeros of the order-parameter field:
\ba
\psi \left[ \vec{r}_i(t),t \right] = 0 .
\ea
Focus on a particular interface, and designate the normal coordinate as
$n$ (with dimensionality 1) and the interfacial coordinates as $\vec{a}$
[with dimensionality $(d-1)$]. Then, the normal velocity $v_n (\vec{a},t)$
obeys the integro-differential equation \cite{to88,lm92}:
\ba
\label{int}
4 \int d \vec{a'} G[\vec{r}_i(\vec{a}), \vec{r}_i(\vec{a'})] v_n (\vec{a'},t)
&\simeq& (1-\alpha) \sigma K (\vec{a},t) + \nonumber \\
&& 4 \alpha \int d \vec{a'}
G[\vec{r}_i (\vec{a}), \vec{r}_i (\vec{a'})] \nabla^2 K (\vec{a'},t) ,
\ea
where $K (\vec{a},t)$ is the local curvature at point $\vec{a}$ on the interface,
and $\sigma$ is the surface tension. The Green's function $G(\vec{x},\vec{x'})$ obeys
\ba
-\nabla^2 G(\vec{x},\vec{x'}) = \delta (\vec{x} - \vec{x'}) .
\ea

A dimensional analysis of Eq.~(\ref{int}) in the scaling regime yields the
growth laws due to surface and bulk diffusion. We identify the scales of
various quantities in Eq.~(\ref{int}) as
\ba
&& [d\vec{a}] \sim R^{d-1}, \quad [G] \sim R^{2-d}, \nonumber \\
&& [v_n] \sim \frac{dR}{dt}, \quad [K] \sim R^{-1}.
\ea
This yields the crossover behavior of the length scale as
\ba
R(t) & \sim & (\alpha t)^{1/4} , \quad t \ll t_c , \nonumber \\
& \sim & [(1-\alpha) \sigma t]^{1/3}, \quad t \gg t_c ,
\ea
where the crossover time is
\ba
\label{tc}
t_c \sim \frac{\alpha^3}{(1-\alpha)^4 \sigma^4} .
\ea

\begin{figure}[htb]
\centering
\includegraphics[width=0.8\textwidth]{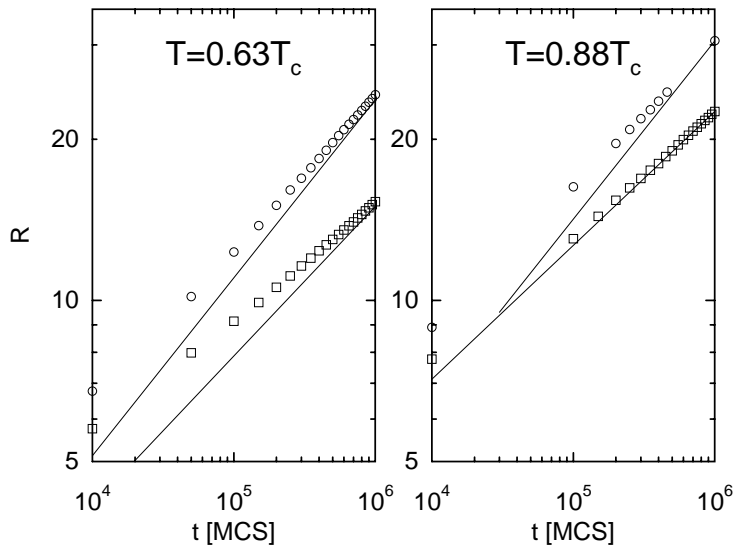}
\caption{Typical domain size ($R$) as a function of time ($t$) after a quench
at $t=0$ from $T=\infty$ to $T=0.63T_c$ (left) and $0.88T_c$ (right). The
circles and squares indicate data obtained with Model B and Model S dynamics,
respectively. Lines with slope $1/4$ and $1/3$ are also provided on the plots
as guides to the eye.
\label{domain}}
\end{figure}

The above scenario applies for both Models B and S, as $D_b < D_s$ in either
case. At moderate temperatures, this crossover occurs rapidly for Model B in our
simulations. However, in Model S, there is a drastic suppression
of bulk diffusion with $D_b \ll D_s$ and $\alpha \simeq 1$. Therefore, the
crossover to $t^{1/3}$-growth
is strongly delayed and not observed over simulation time-scales. This is
seen in Fig.~\ref{domain}, which plots $R$ vs. $t$
at temperatures $T=0.63 T_c$ and $0.88 T_c$ for Models B and S. The data
for Model S is consistent with the growth law $R \sim t^{1/4}$.
The early-time data for Model B is also consistent with this
growth law, as surface diffusion is dominant at early times. At late
times, one sees crossover behavior between the $t^{1/4}$-regime and the
asymptotic $t^{1/3}$-regime. Note that the crossover for Model B is delayed
at the higher temperature $T=0.88 T_c$ because the decrease in $\alpha$ is
more than compensated by the reduction in $\sigma$ due to softening
of the interfaces as $T \rightarrow T_c^-$ [see Eq.~(\ref{tc})].
More generally, we stress that it has
been notoriously difficult to observe the asymptotic $t^{1/3}$-growth in
MC simulations of the Kawasaki model \cite{asm88,mb95}.
Similar results have been obtained from Langevin studies
of coarse-grained models \cite{pbl97,cy89,lm92}.

Before proceeding, we remark that we have also studied models where bulk
diffusion is more strictly suppressed by imposing additional kinetic
constraints which eliminate 2-spin diffusion, 3-spin diffusion, etc. The
corresponding results for the growth law are numerically indistinguishable
from the Model S results in Fig.~\ref{domain} over the time-scales of
our simulation. This underlines the utility of the proposed Model S in
the context of phase separation via surface diffusion.

\begin{figure}[htb]
\centering
\includegraphics[width=0.6\textwidth]{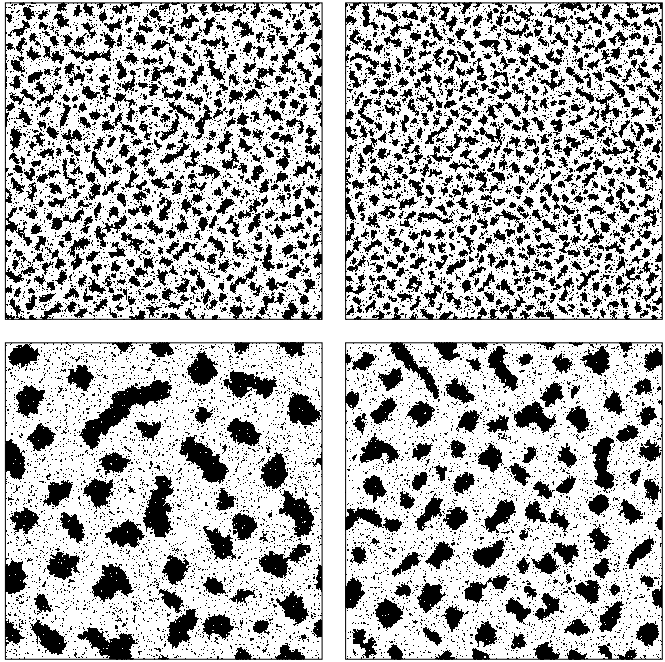}
\vskip1cm
\caption[]{Analogous to Fig.~\ref{snap}, but for an off-critical quench
with 25\% A and 75\% B. The temperature is $T=0.63T_c$.
\label{off}}
\end{figure}

\begin{figure}[htb]
\centering
\includegraphics[width=0.8\textwidth]{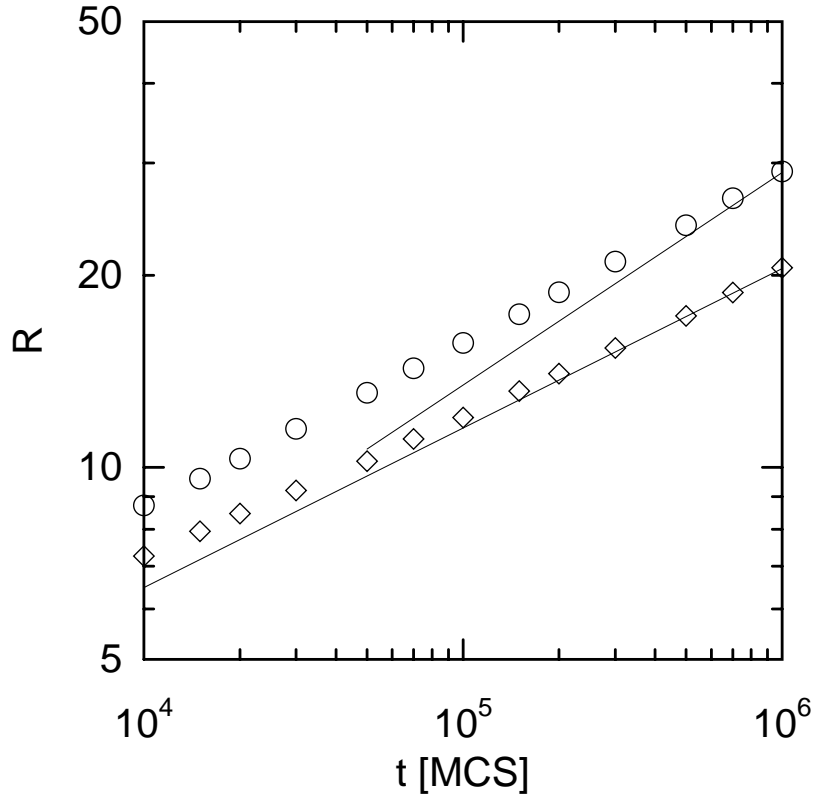}
\vskip0.5cm
\caption[]{Analogous to Fig.~\ref{domain}, but for an off-critical quench
with 25\% A and 75\% B. We show results for $T=0.63 T_c$.
\label{offl}}
\end{figure}

It is also relevant to discuss off-critical quenches, where one of the
components is present in a larger fraction. In Fig.~\ref{off}, we show
evolution pictures for Models B and S for the case with 25\% A and 75\% B.
If the evolution morphology
is not bicontinuous, e.g., there are droplets of the
minority phase in a matrix of the majority phase, the surface-diffusion
mechanism is unable to drive growth. Nevertheless, at high temperatures, growth
may still proceed by the Brownian motion of droplets \cite{hf84}. The 
corresponding growth law depends explicitly on the dimensionality:
\ba
R(t) \sim (Tt)^{1/(d+2)} .
\ea
Thus, domain growth through droplet motion obeys the law
$R(t) \sim (Tt)^{1/4}$ in $d=2$, which is analogous to the surface-diffusion
growth law. The growth kinetics of Models B and S for off-critical mixtures at
$T=0.63 T_c$ is shown in Fig.~\ref{offl}. We see that the S-dynamics shows the
expected $t^{1/4}$-growth over extended time-regimes. As before, the B-dynamics
shows a crossover behavior between the $t^{1/4}$-regime and the asymptotic
$t^{1/3}$-regime. At low temperatures, the Brownian
mechanism is ineffective and the evolution of Model S freezes into a
meso-structure.

\subsection{Correlation Functions}

Next, let us study the morphological features of the evolution in
Figs.~\ref{snap} and \ref{off}. These are usually characterized by (a) the
correlation function defined in Eq.~(\ref{cor}), or (b) its
Fourier transform, the structure factor. We have confirmed that these
quantities exhibit dynamical scaling for both Models B and S. For the
sake of brevity, we do not show these results here.

\begin{figure}[htb]
\centering
\includegraphics[width=0.45\textwidth]{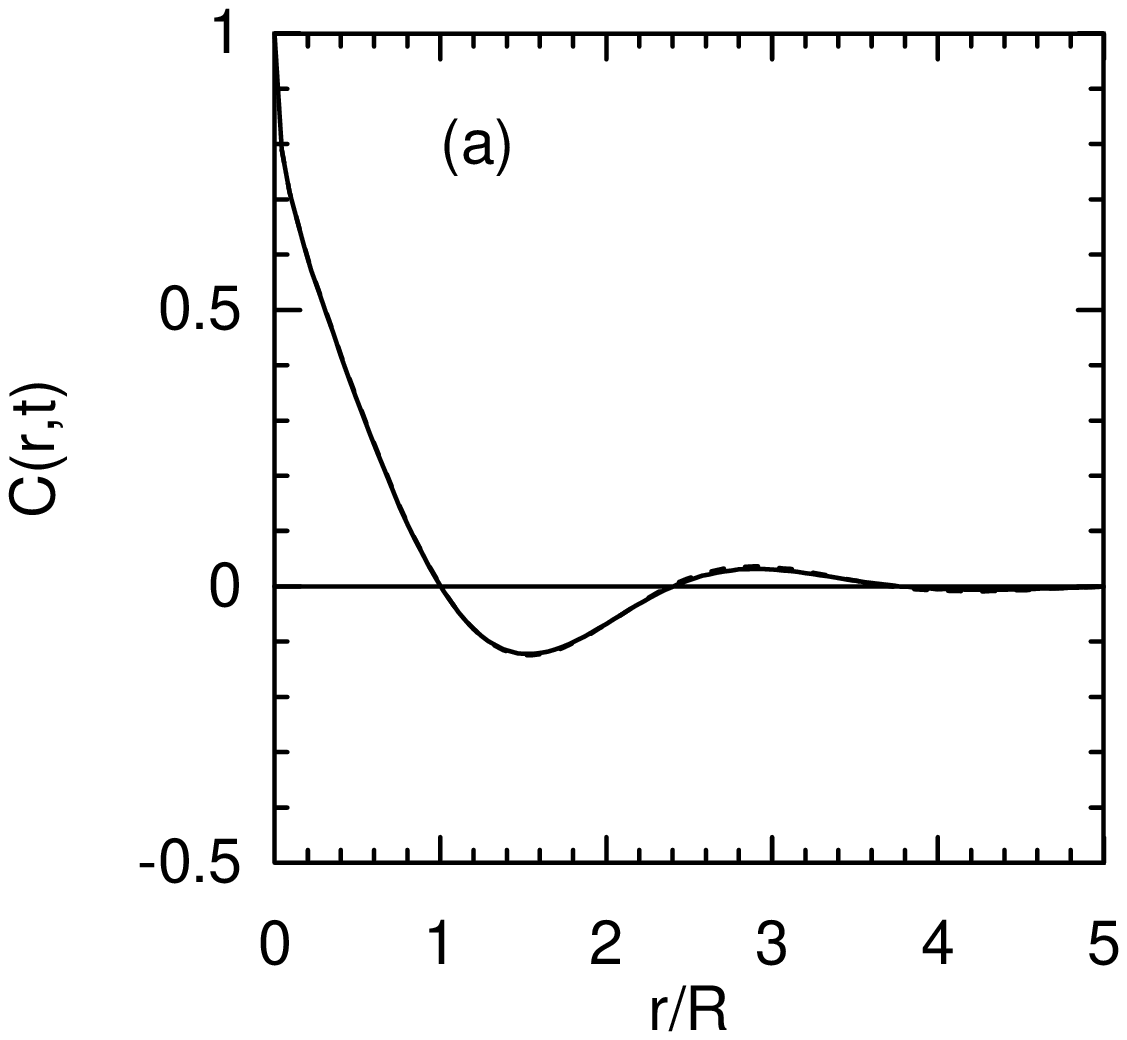}
\includegraphics[width=0.45\textwidth]{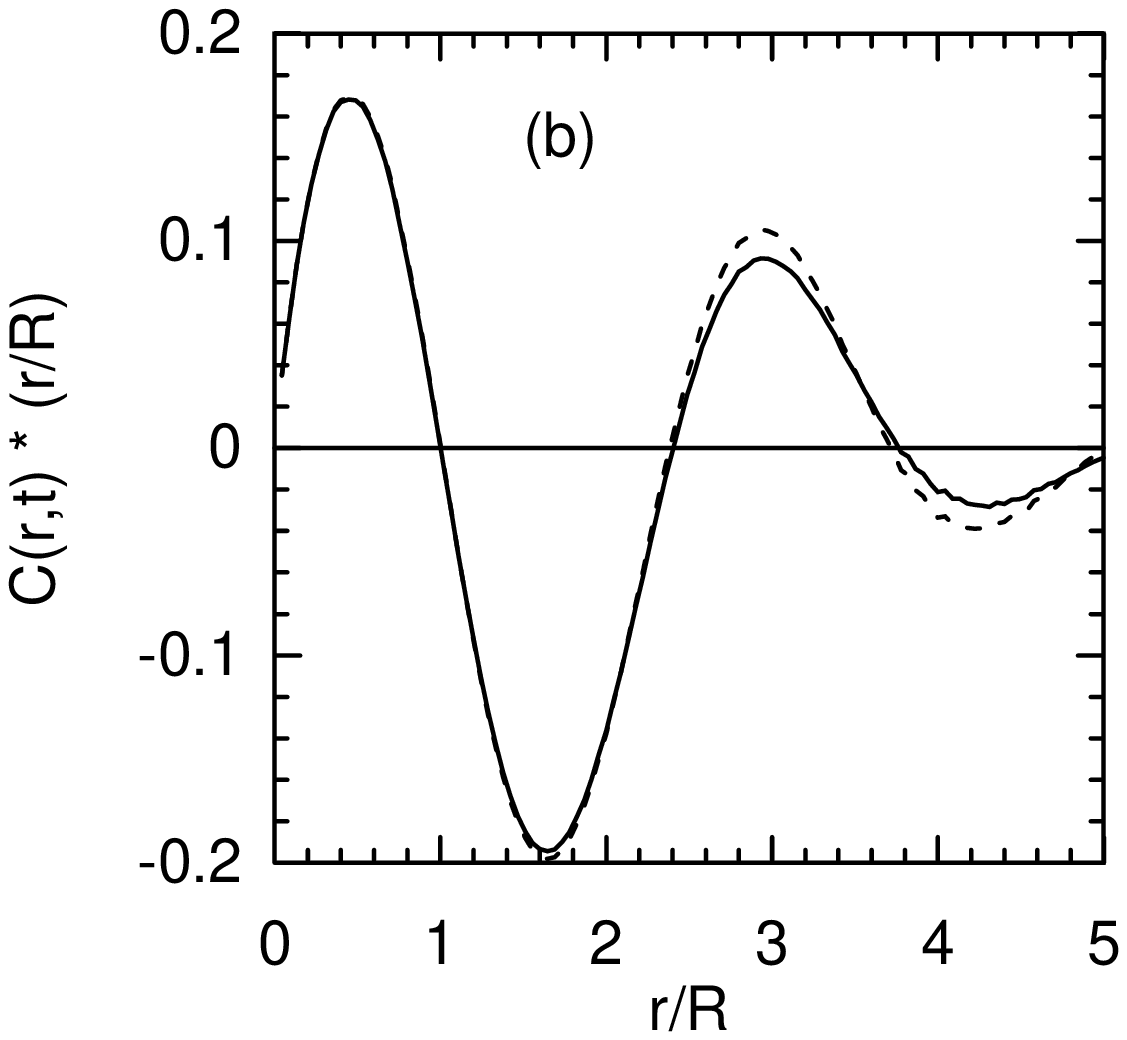}
\caption{Superposition of scaling functions for Models B (solid line)
and S (dashed line) for the evolution depicted in Fig.~\ref{snap}. For Model S,
the data set corresponds to $t=10^6$ MCS; for Model B, the data set corresponds to
$t=3.4 \times 10^5$ MCS. Both domain sizes coincide at these times.
(a) Plot of $C(r,t)$ vs. $r/R$. (b) Plot of $C(r,t) \cdot (r/R)$ vs. $r/R$,
so as to magnify the tail behavior.
\label{grcompare}}
\end{figure}

An important theme in this context is a comparison of the morphologies
arising from both dynamics. Earlier studies with coarse-grained
models \cite{pbl97,cy89,lm92} have found that the correlation functions and
structure factors are numerically indistinguishable for growth driven
by bulk diffusion or surface diffusion. At the visual level, this also
seems to be suggested by the snapshots in Figs.~\ref{snap} and \ref{off}. In
Fig.~\ref{grcompare}, we compare the scaling
functions for Models B and S for a critical quench with $T=0.63 T_c$.
To eliminate finite-size effects, we consider cases with the same
typical domain size: $t=10^6$ MCS in Model S and $t=3.4 \times 10^5$ MCS
in Model B. In Fig.~\ref{grcompare}(a), we plot $C(r,t)$ vs. $r/R$.
The scaling functions superpose on the scale of the plot, in accordance
with earlier studies of phenomenological models. In Fig.~\ref{grcompare}(b),
we plot $C(r,t) \cdot (r/R)$ vs. $r/R$ so that the large-distance behavior
is magnified. Some subtle differences between the two functions are seen at large
distances $r/R > 2$. We make some observations in this regard: \\
(a) The statistical error in the difference between the curves
at the second peak is about four times smaller than the difference,
so it cannot be attributed to statistical fluctuations. \\
(b) We have also replotted the correlation functions for Models B and S
from different times on the scale $C(r,t) \cdot (r/R)$ vs. $r/R$. In that
case, the secondary peaks show a much better collapse, suggesting that the
discrepancy in Fig.~\ref{grcompare}(b) is not the result of corrections
to scaling.

\begin{figure}[htb]
\centering
\includegraphics[width=0.45\textwidth]{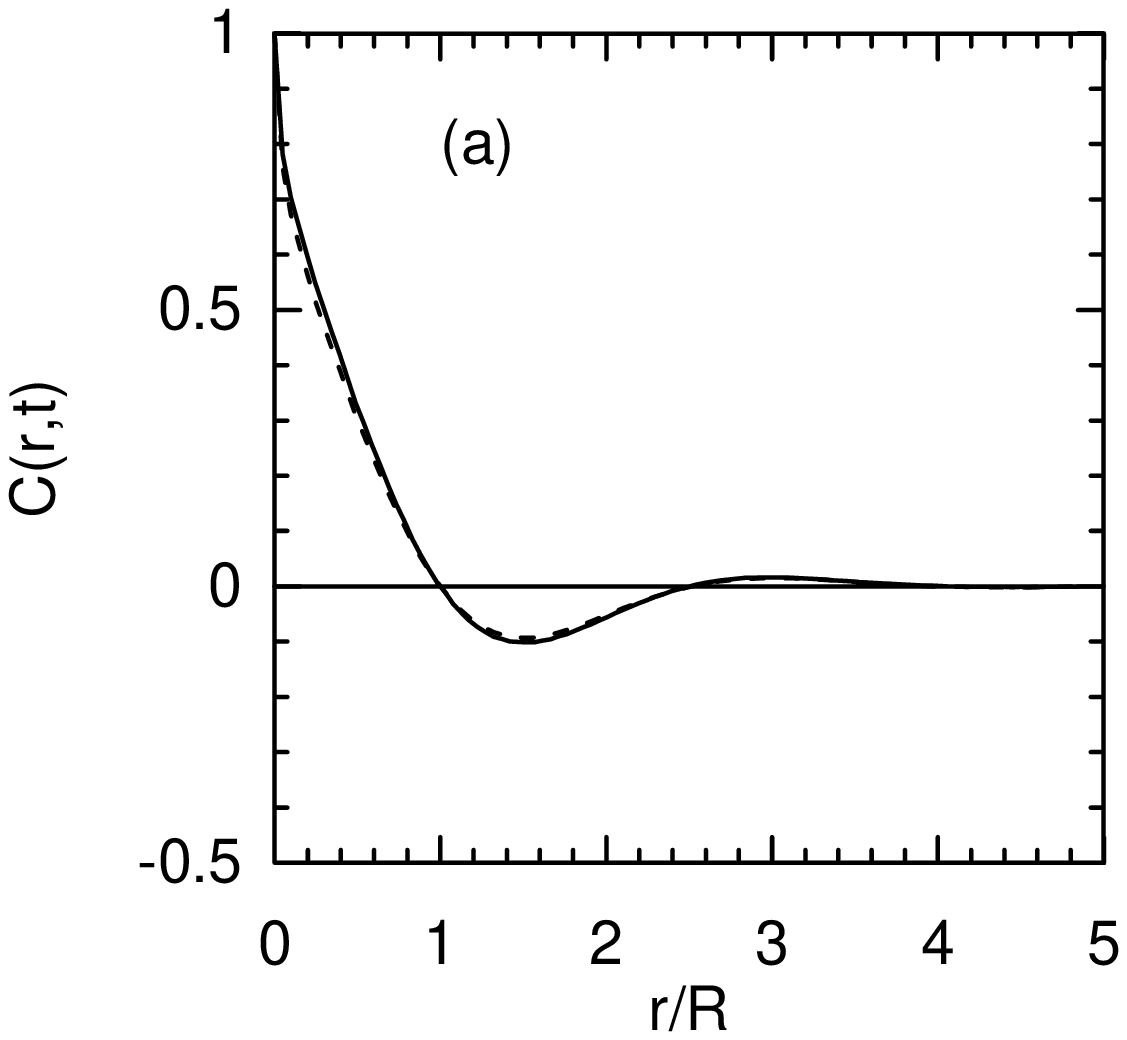}
\includegraphics[width=0.45\textwidth]{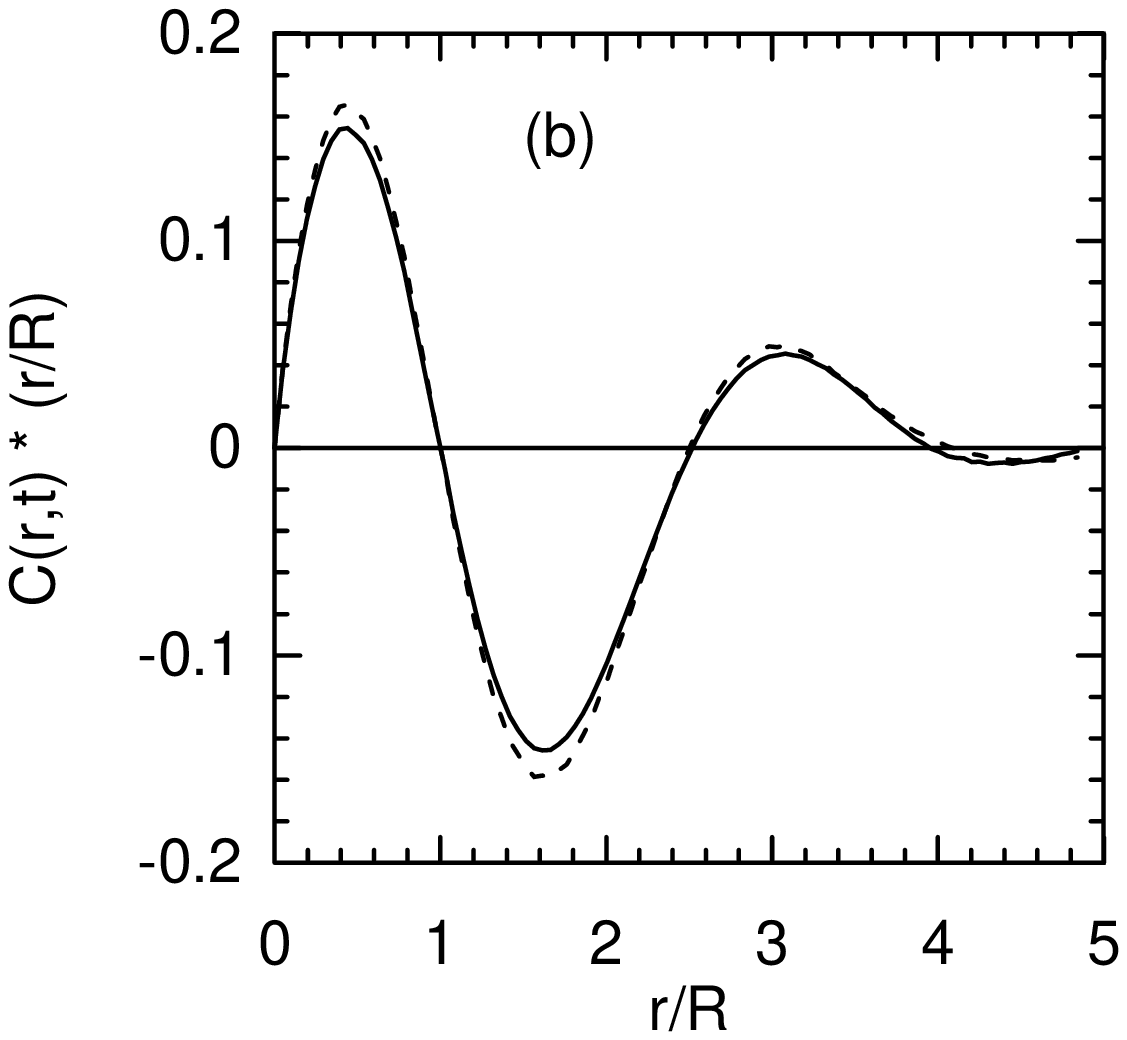}
\caption{Analogous to Fig.~\ref{grcompare}, but for an off-critical quench
with 25\% A and 75\% B at $T=0.63 T_c$. In this case, the times for the different
data sets are $t = 9.8\times 10^5$ MCS (Model S) and $t= 2.7\times 10^5$ MCS
(Model B).
\label{offsc}}
\end{figure}

Though it is difficult to attribute physical significance to these differences,
it is conceptually important to stress the observable differences between the
morphologies for Models B and S. Similar scaling plots for the
off-critical quench shown in Fig.~\ref{off} are shown in Fig.~\ref{offsc}.
It is known that the scaling functions for phase-separating systems depend
on the degree of off-criticality \cite{sp88}. Notice that the oscillations
in the plot of $C(r,t)$ vs. $r/R$ diminish with increase in the off-criticality.
Further, the discrepancy between the
scaling functions for Models B and S is larger for the off-critical case.

\subsection{Island Distribution and Excess Energy}

Our subsequent results will focus on the case of a critical quench.
An alternative method of describing the domain morphology is the
island-size distribution.  We define an island as
a set of aligned spins, all of whose neighbors are either part
of the island, or have an antiparallel spin. Tafa et al. \cite{tpk01} have
shown that the domain-size distribution in a phase-separating system
$\widetilde{P}(l,t)$ exhibits  scaling, and has an exponential decay:
\ba
\label{fx}
\widetilde{P}(l,t) = R^{-1} f\left( \frac{l}{R} \right), \quad
f(x) \sim e^{-ax}~~\mbox{for}~~x \rightarrow \infty ,
\ea
where $l$ is the domain size and $a$ is a constant. The corresponding
scaling form for the island-size distribution $P(s,t)$ in $d=2$ is obtained as
\ba
\label{gx}
P(s,t) &=& \int_0^\infty dl \delta (s-bl^2) \widetilde{P}(l,t) \nonumber \\
&=& \la s \ra^{-1} ~g\left( \frac{s}{\la s \ra} \right) ,
\quad g(x) = \frac{1}{2\sqrt{x}} f(\sqrt{x}) ,
\ea
where $b$ is a geometric factor, and $\la s \ra$ is the average island size.

\begin{figure}[htb]
\centering
\includegraphics[width=0.7\textwidth]{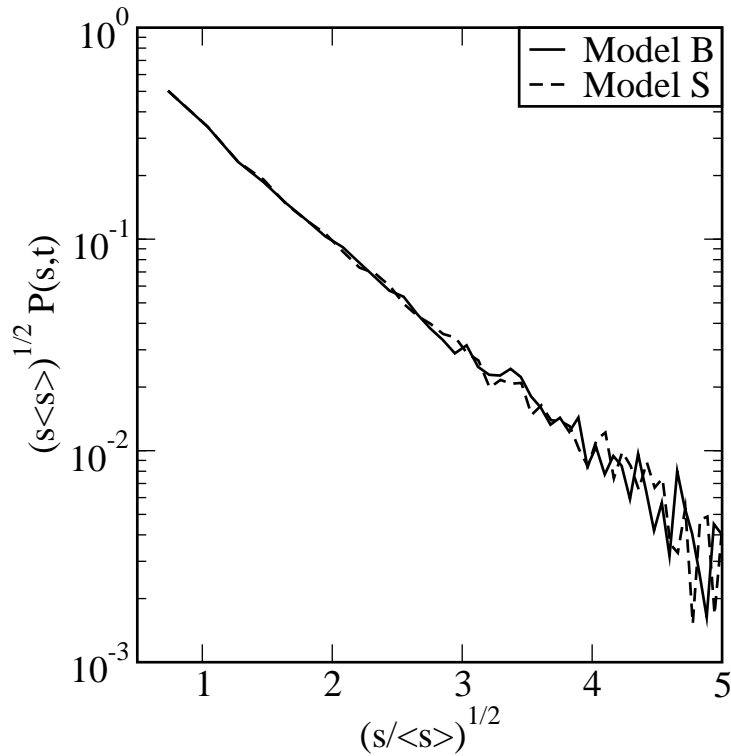}
\caption{Scaled probability distributions for island-sizes in Models B (solid
line) and S (dashed line) at $T=0.88 T_c$. The data is shown on a linear-log
plot as $\sqrt{s \la s \ra} P(s,t)$ vs. $\sqrt{s/\la s \ra}$, suggested by
Eqs.~(\ref{fx})-(\ref{gx}). For Model S, the data set corresponds
to $t=10^6$ MCS; for Model B, the data set corresponds to $t=3.4 \times 10^5$ MCS.
Both domain sizes coincide at these times.
\label{histo}}
\end{figure}

In Fig.~\ref{histo}, we plot
$\sqrt{s \la s \ra} P(s,t)$ vs. $\sqrt{s/\la s \ra}$ for both
Models B and S. We make two observations in this context. First,
the data for the two models is numerically indistinguishable on the
scale of this plot. The subtle differences in the correlation-function
data are not seen in the island-size distribution function.
Second, the plot in Fig.~\ref{histo} exhibits an exponential decay,
as expected from Eqs.~(\ref{fx})-(\ref{gx}).

\begin{figure}[htb]
\centering
\includegraphics[width=0.8\textwidth]{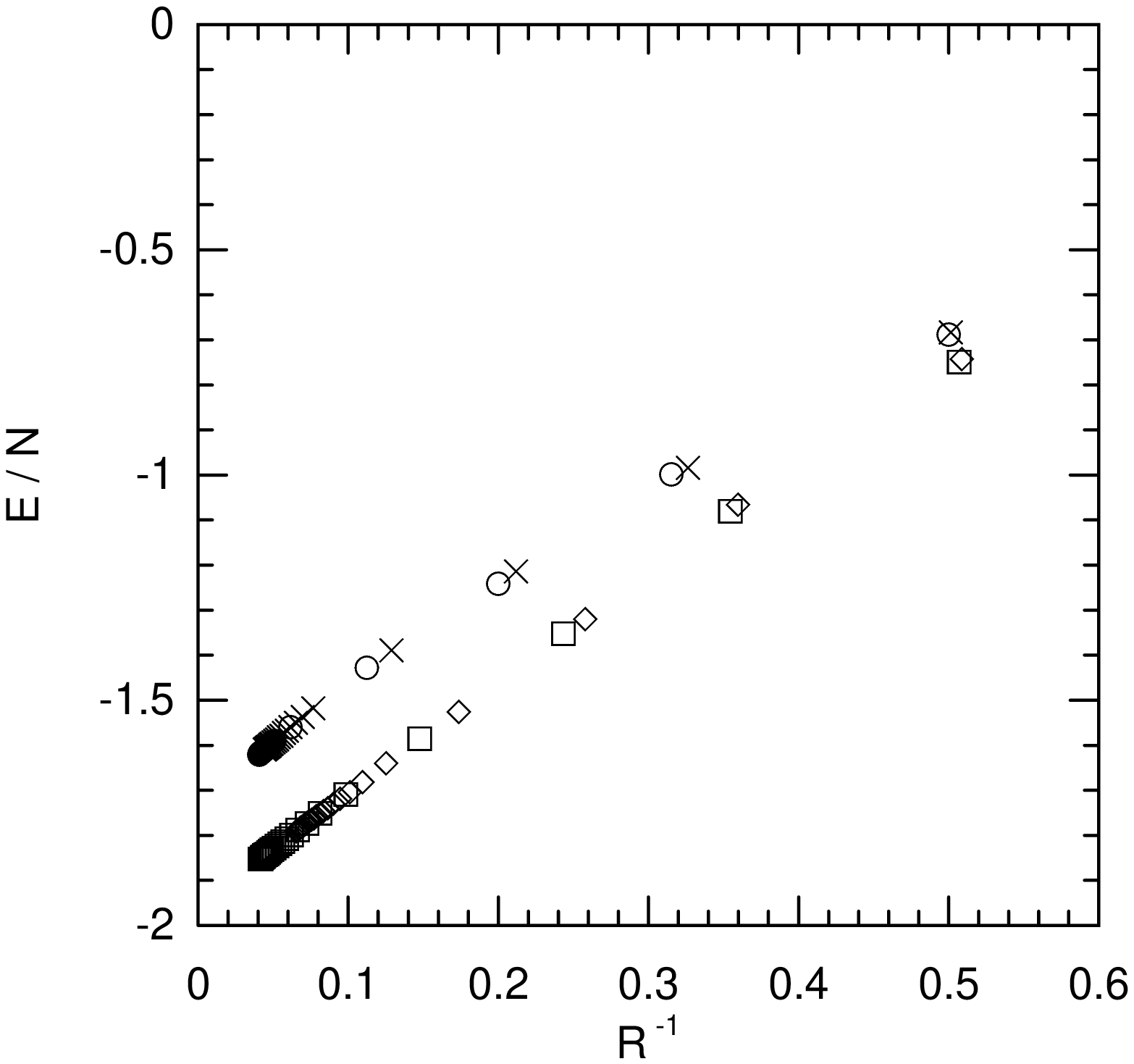}
\caption{Plot of the energy per site $E(t)/N$ vs. $R^{-1}$ for Model B
at temperatures $T=0.63T_c$ (squares) and $T=0.88T_c$ (circles), and
Model S at the same two temperatures (diamonds and crosses).
\label{ener}}
\end{figure}

A macroscopic quantity which depends on the density of small islands is
the total energy $E(t)$. The interfacial energy for a domain is $\sigma R^{d-1}$,
and the number of domains in the system $\sim N/R^d$. Thus, the overall
interfacial energy depends on the length scale as $E(t) - E(\infty)
\sim N\sigma/R$. In Fig.~\ref{ener}, we plot
$E(t)/N$ vs. $R^{-1}$ for both Models B and S at $T=0.63 T_c$ and $0.88 T_c$.
We observe a power-law convergence of the excess energy with
the slope being proportional to the surface tension $\sigma (T)$.
Again, the data sets at the same temperature cannot be distinguished on the
scale of the plot.

\subsection{Aging of the Autocorrelation Function}

The data presented so far has focused on the morphological features
of the phase-separating system. Let us next study the temporal correlation
of the pattern dynamics in Fig.~\ref{snap}. This is measured by the
autocorrelation function:
\begin{eqnarray}
\label{auto}
A(t_w,\tau) & = & {1 \over N} \sum_{i=1}^N \left[ \la \sigma_i (t_w)
\sigma_i (t_w+\tau) \ra  - \la \sigma_i (t_w) \ra \la \sigma_i (t_w+\tau)
\ra \right] ,
\end{eqnarray}
where the times $t_w$ and $(t_w + \tau)$ are measured after the quench at
$t=0$. Here, $t_w$ is the reference time for measurement of the autocorrelation
function, and is referred to as the {\it waiting time}.
The most general correlation function corresponds to unequal space and
time, and combines the definitions in Eqs.~(\ref{cor}) and (\ref{auto}).
Equilibrium systems are stationary and the corresponding $A(t_w,\tau)$ only depends
upon the time-difference $\tau$. On the other hand, for nonequilibrium systems,
$A(t_w,\tau)$ depends on both $t_w$ and $\tau$.

There have been some earlier studies of $A(t_w,\tau)$
for domain growth in kinetic Ising models. There are two
mechanisms which drive the decorrelation process: \\
(a) First, there are fluctuations in bulk domains, which
give a stationary contribution. Huse and Fisher \cite{hf87} studied
decorrelation arising from the appearance of a droplet
of (say) down-spins in an up-domain. The probability that
a droplet of size $R$ appears via fluctuations
is $P_d \propto \exp (-\beta \sigma
R^{d-1})$. The lifetime of this droplet is
$\tau \sim R^{1/\phi}$, where $\phi$ is the growth exponent.
Thus, the corresponding autocorrelation
function shows a stretched-exponential behavior:
\ba
&& A_{\rm{eq}}(\tau) \simeq \exp (- \beta \sigma \tau^\theta), \nonumber \\
&& \theta = (d-1) \phi~\mbox{for}~d<d_c~\mbox{and}~1~\mbox{for}~d>d_c.
\ea
Here, the critical dimensionality is defined by $(d_c-1) \phi = 1$. \\
(b) Second, there is decorrelation due to domain-wall motion. This can be
either stochastic (due to thermal fluctuations) or systematic (due to the
curvature-reduction mechanism). Consider the $T=0$ case, where there are no
fluctuations in the bulk or the surface.
The characteristic domain-wall velocity decreases with time, so this mechanism
gives a non-stationary or aging ($t_w$-dependent) contribution
\cite{bckm97}. Fisher and Huse \cite{fh86} used scaling ideas
to argue that the aging contribution to $A(t_w,\tau)$ has a power-law
dependence on the length scale:
\be
\label{power}
A_{\rm{age}} (t_w,\tau) = \left[ {R(t_w) \over
R(t_w+\tau)} \right]^{\lambda}, \quad R(t_w+\tau) \gg R(t_w) .
\ee
There have been various studies of the aging exponent
$\lambda$ in cases with both spin-flip and
spin-exchange kinetics \cite{ab94}. For power-law domain growth,
Eq.~(\ref{power}) obeys the scaling form $A_{\rm{age}}(t_w,\tau) = h(\tau/t_w)$,
which has been observed in some studies of spin glasses \cite{bckm97}.

\begin{figure}[htb]
\centering
\includegraphics[width=0.8\textwidth]{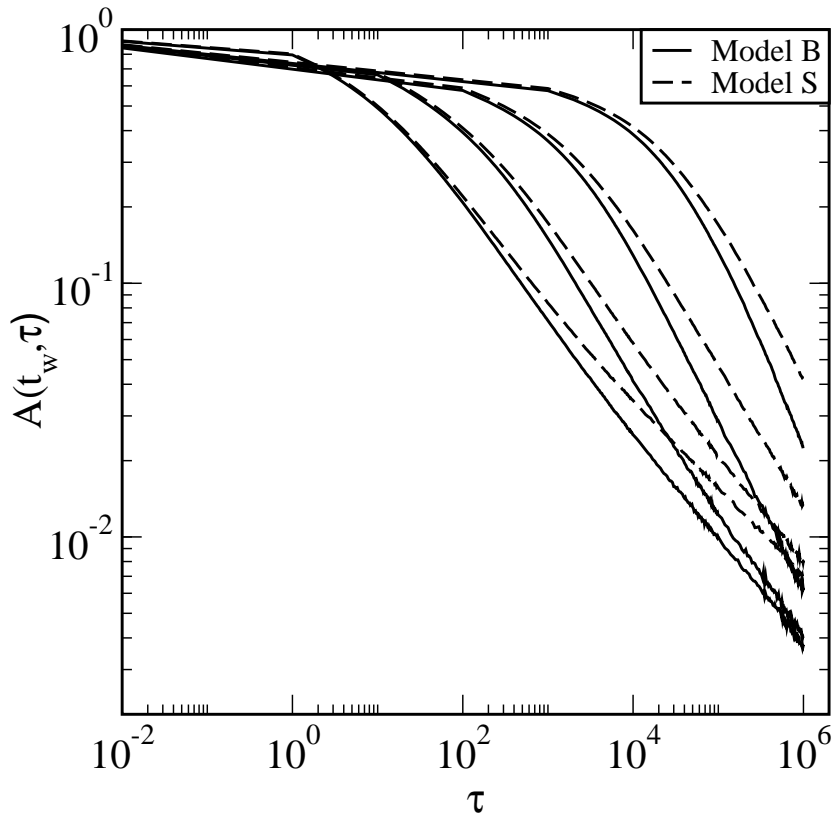}
\caption{Time-dependence of the autocorrelation function for Models B (solid line)
and S (dashed line) at $T=0.63T_c$. The waiting times are $t_w = 10^2, 10^3,
10^4, 10^5$ MCS (from left to right).
\label{age}}
\end{figure}

In Fig.~\ref{age}, we plot $A(t_w,\tau)$ vs. $\tau$ for Models B and S for a
critical quench to
$T=0.63 T_c$. The solid lines denote data for Model B with waiting times
$t_w = 10^2, 10^3, 10^4, 10^5$ (from left to right). The dashed lines denote the
corresponding data for Model S. As expected, the autocorrelation function decays
more rapidly for Model B. We make the following observations in this regard: \\
(a) The quantity $A(t_w,\tau)$ exhibits aging, with an explicit dependence on $t_w$
for both Models B and S. In general, the decay is slower for larger $t_w$, i.e.,
when the domain size of the reference state is larger.
Further, the decay is faster for higher temperatures,
where larger fluctuations are present in the system. \\
(b) The data in Fig.~\ref{age} is plotted on a log-log scale,
and exhibits a continuous curvature for both Models B and S. This is not
consistent with the simple power-law decay in Eq.~(\ref{power}). As a matter
of fact, the autocorrelation data does not even exhibit $\tau/t_w$-scaling,
as we have confirmed. In the case of
Model B, this is because the decorrelation process is driven by both bulk fluctuations
(with a stationary contribution) and domain-wall motion (with a
non-stationary contribution). In the case of Model S, bulk fluctuations
have been effectively eliminated and one may naively expect to recover
power-law decay. However, this is not the case because interfacial
fluctuations also contribute to decorrelation. We believe that the scaling
behavior in Eq.~(\ref{power}) is only realized in kinetic Ising models or their
coarse-grained analogs at $T=0$. In this limit, coarsening occurs only through
the systematic motion of interfaces and the system always reduces its energy.
However, the $T=0$ limit is not interesting in the context of kinetic Ising
models because the evolving system invariably gets trapped in local free-energy
minima. \\
(c) In recent work, Puri and Kumar \cite{pk04} have studied the decorrelation
process in a spin-1 model using a stochastic model based on the continuous-time
random walk formalism. We are currently trying to adapt their modeling
to understand the behavior in Fig.~\ref{age}.

\section{Summary and Discussion}

Let us conclude this paper with a summary and discussion of the results presented
here. We have studied phase separation in a kinetic Ising model for phase
separation mediated by surface diffusion. This model (referred to as Model S)
is obtained by imposing a kinetic constraint on the usual Kawasaki kinetic
Ising model (referred to as Model B). In general, the surface diffusion
mechanism can drive segregation only when the morphology consists of percolated
clusters, i.e., for near-critical quenches. We have undertaken Monte Carlo (MC)
simulations of Models B and S using
multi-spin coding techniques. These provide accelerated algorithms which
enable the simulation of large systems for extended times. Our results
show that the major difference between the morphologies of Models B and S
lies in the growth dynamics. In this regard, it is relevant to emphasize
the following: \\
(a) The early-time dynamics ($t \ll t_c^B$) of Model B is also dominated by
surface diffusion with the growth law $R(t) \sim t^{1/4}$. For late times
($t \gg t_c^B$), there is a crossover to the $t^{1/3}$-growth regime.
In the limit of $T \rightarrow 0$, the crossover time diverges
($t_c^B \rightarrow \infty$).
However, the low-temperature dynamics of Model B usually
freezes into metastable states. Therefore, it is hard to see an extended
regime of $t^{1/4}$-growth in Model B. \\
(b) Our kinetic constraint eliminates single-particle bulk diffusion,
and we see extended regimes of growth driven by surface diffusion. However,
$n$-particle diffusion (with $n \geq 2$) is still possible and is governed
by the probability for existence of impurities in bulk domains. Thus, at
sufficiently large times ($t \gg t_c^S$), we again expect a crossover to
$t^{1/3}$-growth. However, this crossover is extremely delayed,
even at moderate temperatures. \\
(c) We have also studied kinetic models with constraints which
eliminate the diffusion of $n$-spin clusters. The domain growth data obtained
from these models is numerically indistinguishable from that for Model S. \\
(d) For highly off-critical quenches, the morphology consists of droplets
of the minority phase in a matrix of the majority phase. In this case, the
surface diffusion mechanism cannot drive phase separation. However, Brownian
motion and coalescence of droplets also gives rise to $t^{1/4}$-growth in $d=2$.

Apart from growth laws, we have also studied quantitative properties
of the evolution morphology like correlation functions and island-size
distribution functions. There are subtle differences in the scaled
correlation functions for Models B and S, but it is difficult to attribute
physical significance to these. Further, these differences are not reflected
in the island-size distribution function.

Finally, we have studied the aging of the autocorrelation function $A(t_w,\tau)$
in Models B and S. In both cases, we find that the decorrelation process is
driven by both fluctuations and domain-wall motion. Thus, $A(t_w,\tau)$ does
not exhibit a simple power-law decay or scaling behavior. We are presently
adapting the continuous-time random walk approach developed by
Puri and Kumar \cite{pk04} to study the aging of the autocorrelation functions
in Models B and S.

\end{document}